\documentclass[12pt,preprint]{aastex}

\begin{document}

\title{H$\beta$ Line Widths as an Orientation Indicator for Low-Ionization Broad Absorption Line Quasars}
\author{Brian Punsly\altaffilmark{1}} \and\author{Shaohua Zhang\altaffilmark{2}}
\altaffiltext{1}{4014 Emerald Street No.116, Torrance CA, USA 90503
and ICRANet, Piazza della Repubblica 10 Pescara 65100, Italy,
brian.punsly@verizon.net or
brian.punsly@comdev-usa.com}\altaffiltext{2}{Key Laboratory for
Research in Galaxies and Cosmology, University of Sciences and
Technology of China, China Academy of Science, Hefei, Anhui, 230026
China}

\begin{abstract}
There is evidence from radio-loud quasars to suggest that the distribution
of the H$\beta$ broad emission line (BEL) gas is arranged in a predominantly
planar orientation, and this result may well also apply to radio-quiet quasars.
This would imply that the observed full
width at half maximum (FWHM) of the H$\beta$ BELs is dependent on
the orientation of the line of sight to the gas. If this view is
correct then we propose that the FWHM can be used as a surrogate, in
large samples, to determine the line of sight to the H$\beta$ BELs
in broad absorption line quasars (BALQSOs). The existence of broad
UV absorption lines (BALs) means that the line of sight to BALQSOs
must also pass through the BAL out-flowing gas. It is determined
that there is a statistically significant excess of narrow line
profiles in the SDSS DR7 archival spectra of low ionization broad
absorption line quasars (LoBALQSOs), indicating that BAL gas flowing
close to the equatorial plane does not commonly occur in these
sources. We also find that the data is not well represented by
random lines of sight to the BAL gas. Our best fit indicates two
classes of LoBALQSOs, the majority ($\approx 2/3$) are polar
outflows, that are responsible for the enhanced frequency of narrow
line profiles, and the remainder are equatorial outflows. We further
motivated the line of sight explanation of the narrow line excess in
LoBALQSOs by considering the notion that the skewed distribution of
line profiles is driven by an elevated Eddington ratio in BALQSOs.
We constructed a variety of control samples comprised of
nonLoBALQSOs matched to a de-reddened LoBALQSO sample in redshift,
luminosity, black hole mass and Eddington ratio. It is demonstrated
that the excess of narrow profiles persists within the LoBALQSO
sample relative to each of the control samples with no reduction of
the statistical significance. Thus, we eliminate the possibility
that the excess narrow lines seen in the LoBALQSOs arise from an
enhanced Eddington ratio.
\end{abstract}

\keywords{(galaxies:) quasars: absorption lines --- galaxies: jets
--- (galaxies:) quasars: general --- accretion, accretion disks --- black hole physics}

\section{Introduction}About 15\% - 20\% of quasars show broad UV absorption lines
(originally defined as absorbing gas that is blue shifted at least
5,000 km/s relative to the QSO rest frame and displaying a spread in
velocity of at least 2,000 km/s)
\citep{wey97,hew03,rei03,tru06,gib09}. Although evolutionary
processes might be related to BAL outflows this does not seem to be
the primary determinant. There is no indication that there is an
excessive amount of radiation associated with reprocessed emission
from dust as would be expected if the BALQSOs were in an
evolutionary stage characterized by a large volume of dusty gas
enshrouding the central AGN \citep{wil03, gal07}. Also, the overall
optical/UV spectra are strikingly similar to non-BALQSOs
\citep{wey91,rei03}. Thus, it is widely believed that all or most
radio quiet quasars have BAL flows, but the designation of a quasar
as a BALQSO depends on whether the line of sight intersects the
solid angle subtended by the outflow. This prevailing view is our
fundamental assumption that motivates our search for a diagnostic
that determines these preferred lines of sight. This is our primary working
hypothesis and we explore possible effects that would be expected by preferred lines
of sight.
\par The "standard model of quasars" is one of a
hot accretion flow onto a black hole and a surrounding torus of
molecular gas \citep{ant93}. Theoretical treatments indicate that
the BAL outflow can be an equatorial wind driven from the outer
regions of a luminous accretion disk that is viewed at low
latitudes, \citet{mur95}, or a bipolar flow launched from the inner
regions of the accretion flow \citep{pun99,pun00}. Furthermore,
\citet{elv00}, proposed a purely phenomenological model in which the
BAL wind begins near or at the accretion disk and flows at
mid-range latitudes arranged so as to propagate just above the
surface of the distant dusty torus. It has also been proposed that
there is more than one wind source (polar and equatorial) for the
BAL winds that coexist in QSOs \citep{pun00,pro04}. It has been further
argued by \citet{bro06} that observations rule out one preferred
narrow range of lines of sight to the BAL wind. The question is left
open as to the distribution of lines of sights to BALQSOs, a
question we hope to answer for LoBALQSOs in this study.
\par There is very little direct evidence that we have on the line of sight to the BAL region.
The most direct method implemented so far is to use radio
variability information \citep{zho06,gho07}. This information can be
used to bound the size of the radio emitting gas then deduce that
the radio emission must be viewed close to the polar axis and
emanate from a relativistic jet, thereby avoiding the well known
inverse Compton catastrophe \citep{mar79,kel69,lin85}. However, this
method will only find sources with a polar orientation and it is
limited to a subset of those sources that have sufficient radio flux
to make the measurement and multiple epochs of observation with
compatible sensitivity, a very small subsample. Thus, another
orientation indicator is needed in this field of research. In radio
loud quasars, the dominance of the radio core relative to the large
scale radio lobes is the standard orientation indicator
\citep{bro86}. Core dominant objects are polar, lobe dominant
objects are viewed at large angles from the jet axis. Almost all
BALQSOs are unresolved with the VLA, \citet{bec00}, but the radio
fluxes are small and there is not sufficient sensitivity (and in
many cases insufficient resolution) to get a meaningful estimate of
core dominance. Furthermore, the spectra of BALQSOs are often similar to
a special class of radio source, Gigahertz peaked radio sources, for
which this orientation indicator might not even be applicable
\citep{mon08}.
\par A breakthrough paper in the field of BEL orientation was
\citet{bro86} which showed that there was an anti-correlation
between radio core dominance and the FWHM of H$\beta$ in radio loud
quasars that was explained by a broad line region (BLR) that was
predominantly planar. The existence of a planar distribution of low
ionization BEL gas seems to carry over to the radio quiet regime
as well. It was noted in \citet{ant89,mai02} that a planar
distribution resolves the paradox that quasar Ly$\alpha$ equivalent
widths (and BEL equivalent widths in general) imply column densities
that should show the Lyman continuum in absorption and with all
profiles heavily damped. But, no quasar BEL cloud has ever been
convincingly seen in absorption \citep{ant89}. Since a large
fraction of the solid angle around quasar must be covered with
low-ionization BEL gas, yet no Lyman continuum absorption has ever
been convincingly detected from the BEL clouds, it must be that
quasars oriented favorably for BEL absorption are not seen as
quasars.  The only plausible way to do this is by placing the low
ionization BEL clouds within the solid angle of the torus, so that
these objects are not seen as quasars from our point of view.
Furthermore, the large covering factor of the continuum by the
equatorial planar BEL gas distribution creates an enormous volume of
ionized BEL gas, sufficient to produce these large equivalent
widths. Lines of sight far above the equatorial plane (in the hole
of the dusty torus, as expected for exposed quasar nuclei,
\citet{ant93}) penetrate the relatively thin column densities
orthogonal to the plane of the gas distribution, revealing the
continuum in emission, with insufficient column density to produce
damped Lyman absorption edges. This resolves the paradox. In
conclusion, the notion of a planar distribution of low ionization
BEL gas appears to be applicable to all quasars and suggests that
the main ideas of \citet{bro86} are relevant to predominantly radio
quiet samples as well. There is no evidence of a difference between
radio quiet and radio loud AGN in this regard. There are some
differences in the spectra of radio loud and radio quiet quasars
that have been noted in the literature that are not entirely
understood (see eg. \citet{cor94,cor96}), but do not clearly have a
bearing on this point
\par Thusly motivated, we compiled the FWHM of
low redshift, $0.4 < z <0.8$, LoBALQSOs in the SDSS DR7 archives to
look for orientation information. This redshift range insures that
both Mg II and H$\beta$ emission lines are visible. High ionization,
CIV, absorption will not appear in these spectra thus, our analysis
is limited to LoBALQSOs. First we compare the quasars showing LoBALs
with those that do not. Then we simulate the distribution of
H$\beta$ FWHM that would be expected for different lines of sight.
This information is used to discuss the line of sight to the BLR of
LoBALQSOs.
\section{The Distribution of H$\beta$ in SDSS DR7}In the DR7 release we
found 102 LoBALQSOs (out of a sample of 10,069 quasars) in the
redshift range, $0.4 < z <0.8$. The LoBALQSO determination is based
on spectra with S/N $>7$, using the criteria of continuous
absorption over a velocity interval greater than 1600 km/s for a
depth of at least $10\%$ as in \citet{zha10} which see for a
detailed discussion of this criteria. The fitting of the H$\beta$
profile also follows the DR5 analysis of \citet{zha10}. The
distributions of H$\beta$ FWHM of LoBALQSOs and non-LoBALQSOs from
DR7 in this redshift range are compared in Figure 1. Because the bin
populations of the LoBALQSOs are small, error bars were added to the
bin frequencies. The horizontal error bars represent the width of
the bins in Figure 1. The vertical (bin population) error bars are
difficult to assess for small bin sizes, we chose to represent these
with binomial statistics.
\begin{figure}
\includegraphics[width=160 mm, angle= 0]{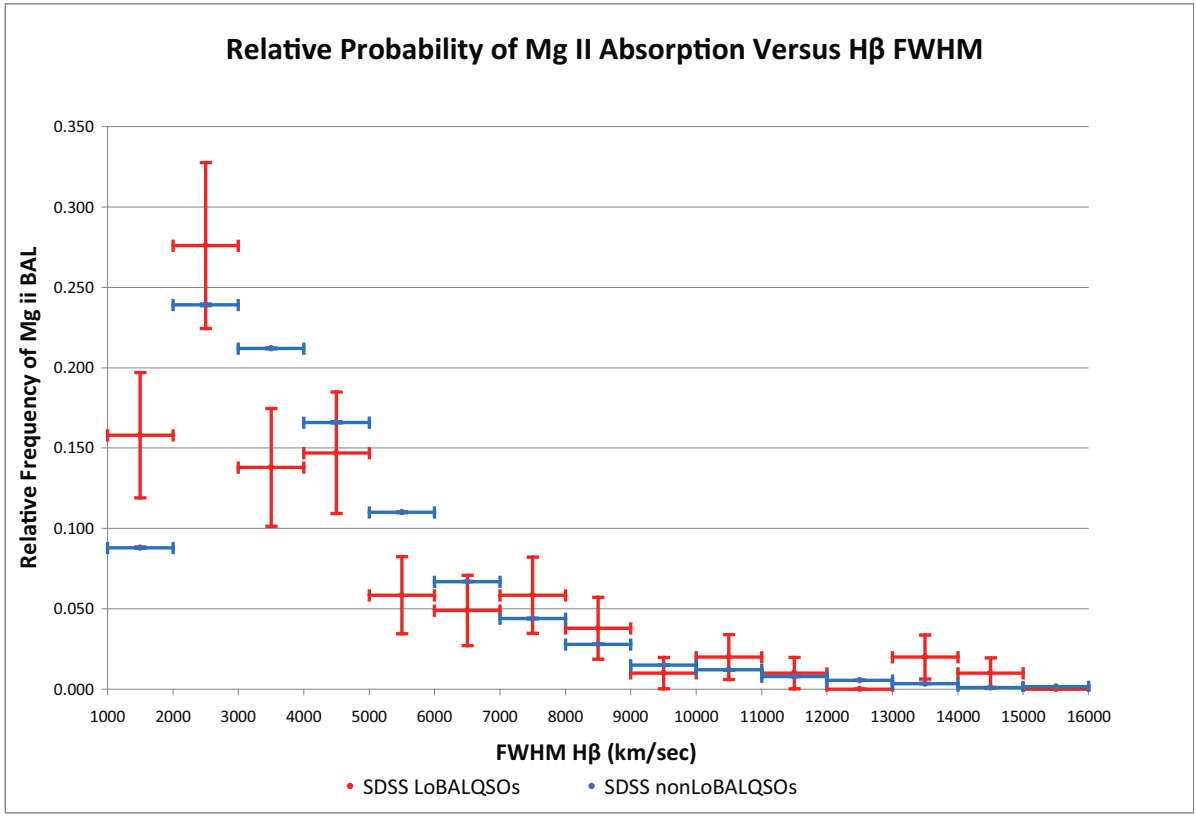}
\caption{Comparison of the distributions of H$\beta$ FWHM of
LoBALQSOs and non-LoBALQSOs from DR7. Note the excess of narrow line
profiles less than 2,500 km/s in the LoBALQSO population. The error
bars are based on binomial statistics.}
\end{figure}
It is notable from Figure 1 that there is a difference in the partial
distributions of FWHM between the two samples for values smaller
than 6,500 km/sec (a subsample that represents $\approx 80\%$ of the
sources for both the total sample and the LoBALQSOs, separately). The
LoBALQSOs are skewed toward lower FWHM in the partial distribution.
The difference between these partial distributions is significant in both a
K-S test (a null probability of 0.0178) and a Wilcoxon Rank Sum Test
(a null probability of 0.0028). Thus, formally, we conclude that the
distribution of LoBALQSO FWHM is poorly described by the
distribution of FWHM of non-BALQSOs. To understand more than this, we
consider the effects of a preferred line of sight to the BAL region
and an elevated Eddington ratio in LoBALQSOs.
\section{Searching for an Alternative Physically Based Explanation}
Before adopting a line of sight interpretation of the excess of
narrow emission lines in the LoBALQSO population in Figure 1, we
explore the possibility that the effect is driven by a physical
difference in the black hole accretion system between the LoBALQSOs
and other quasars. We explore three potential physical parameters:
the bolometric luminosity, $L_{bol}$, the mass of the central black
hole, $M_{bh}$ and the Eddington ratio, $\Gamma= L_{bol}/L_{Edd}$.
To do this, we create "matched subsamples" of the SDSS DR7
nonLoBALQSOs that are matched in optical luminosity, black hole mass
and Eddington ratio to LoBALQSOs. It is demonstrated that the
difference in FWHM between LoBALQSOs and nonLoBALQSOs that was noted
in Figure 1 is just as pronounced in the "matched" nonLoBALQSO
samples. Obviously, all the quasar properties can not be identical
between two samples in all moments of the probability distributions.
One must choose the most important parameter and optimize for that
one. We pay particularly close attention to subsamples that are
matched with respect to $\Gamma$.
\par In our opinion, $L_{optical}=\lambda L_{\lambda}(5100
\AA)$ is the most logical parameter to restrict if one is to
construct matched subsamples for the following reasons:
\begin{enumerate}
\item $L_{optical}$ is directly measurable and the other quantities are derived in consort
with many assumptions.

\item In \citet{zha10}, Figure 6, it was shown that a comparison of 2MASS fluxes with SDSS fluxes that
the rest frame near IR/optical colors are similarly distributed for
LoBALQSOs and nonLoBALQSOs. Thus, this optically selected sample of
LoBALQSO spectra show minimal reddening in the optical band, with a
magnitude that is similar to that of the background galactic
starlight contribution \citep{zha10}. We expect that $L_{optical}$
is representative of the intrinsic optical luminosity up to a small
correction on the order of the galactic starlight. With these small
de-reddening corrections, $L_{optical}$ is an acceptable surrogate
for $L_{bol}$ in either the LoBALQSO sample or nonLoBALQSO
sample.\footnote{Note that if one were to cull the sample of
LoBALQSOs from an IR flux limited sample, we would not expect this
to be true. The SDSS optical selection criteria will not find the
more reddened LoBALQSOs that would be found in the IR samples. We
also note that \citet{zha10} showed that the LoBALQSO spectra are
significantly reddened in the ultraviolet and therefore
$L_{UV}=\lambda L_{\lambda}(3000 \AA)$ is not a good surrogate for
$L_{bol}$}

\item It is shown in \citet{gan07} for HiBALQSOs and corroborated for LoBALQSOs in Figure 12 of \citet{zha10} that the biggest physical difference
between BALQSOs and nonBALQSOs is that BALQSOs tend to have larger
rest frame continuum fluxes.
\end{enumerate}
The significant difference in the rest frame continuum flux noted in
point 3 above, is driven largely by the lack of low values of
$L_{optical}$ within the BALQSO population. This is expected by
dynamical considerations, low values of radiation pressure do not
produce enough force to drive outflows from the environs of a black
hole \citep{mur95,pun00}. Figure 12 of \citet{zha10} indicates that
there are very few DR5 LoBALQSOs with $L_{optical}< 4 \times
10^{44}$ ergs/sec. Thus, we construct matched samples by truncating
the nonLoBALQSO DR7 sample on the low side. We create a variety of
samples by varying the low end cutoff. We produced many such samples
with a range of the derivable physical parameters designed to
straddle those of the LoBALQSOs. Table 1 shows data for the various
samples created with different cutoffs.

\begin{table}
\caption{Physical Properties of Samples}
\footnotesize{\begin{tabular}{ccccccccc} \tableline \rule{0mm}{3mm}
 Sample  & cutoff & Number & $L_{optical}$ & P & P & $\Gamma $ & Log($M_{bh}$)& OIII EW \\
               & $10^{44}$ \tablenotemark{c} & & $10^{45}$\tablenotemark{c} & Wilcoxon& K-S & & $M_{\odot}$& $\AA$  \\
\tableline \rule{0mm}{3mm}
LBQSO\tablenotemark{a} &  0 & 102 & $ 1.05 \pm 1.04 $ & 1 & 1 &$ 0.274 \pm 0.247 $&$ 8.51 \pm 0.60 $& $19.0 \pm 29.4$ \\
LBQSO\tablenotemark{a}   &  0 &  102 & $ 1.32 \pm 1.31 $ & 1 & 1 &$ 0.307 \pm 0.276$ & $ 8.56 \pm 0.60 $& $19.0 \pm 29.4$ \\
(de-reddened) &   &   &  &  & & & &\\
\tableline{\rule{0mm}{3mm}}
NLBQSO\tablenotemark{b} & 0  & 10069 & $ 0.67 \pm 0.75 $ & 0.0028 &0.0178&$ 0.214 \pm 0.223 $&$ 8.44 \pm 0.46 $& $ 26.7 \pm 31.0 $  \\
NLBQSO\tablenotemark{b}  & 5 & 4824 & $ 1.03 \pm 0.95 $ & 0.0008 &0.0054&$ 0.262 \pm 0.261$& $ 8.57 \pm 0.44 $& $ 23.5 \pm 21.3$ \\
NLBQSO\tablenotemark{b}  & 6 & 3706 & $ 1.18 \pm 1.04 $ & 0.0007 & 0.0051 &$ 0.277 \pm 0.271$& $ 8.61 \pm 0.44 $& $ 23.2 \pm 21.0 $ \\
NLBQSO\tablenotemark{b}  & 7  & 2919 & $ 1.32 \pm 1.13 $ & 0.0005 & 0.0049 &$ 0.290 \pm 0.282$& $ 8.64 \pm 0.44 $ & $ 22.7 \pm 20.4 $ \\
NLBQSO\tablenotemark{b}  & 8  & 2313 & $ 1.47 \pm 1.22 $ & 0.0006 & 0.0061 &$ 0.304 \pm 0.293$& $ 8.67 \pm 0.44 $ & $ 22.3 \pm 20.1 $ \\
NLBQSO\tablenotemark{b}  & 9  & 1860 & $ 1.62 \pm 1.32 $ & 0.0005 & 0.0055 &$ 0.316 \pm 0.300$& $ 8.70 \pm 0.43 $ & $ 21.8 \pm 19.6$ \\
NLBQSO\tablenotemark{b}  & 10  & 1478 & $ 1.79 \pm 1.43 $ & 0.0005 & 0.0078 &$ 0.332 \pm 0.313$&$ 8.72 \pm 0.43 $ & $ 21.4 \pm 19.3$ \\
NLBQSO\tablenotemark{b}  & 9\tablenotemark{d}  & 1551 & $ 1.45 \pm 0.87 $ & 0.0017 & 0.0152 &$ 0.321 \pm 0.300$&$ 8.66 \pm 0.42 $ & $ 16.3 \pm 9.0$ \\
NLBQSO\tablenotemark{b}  & 9\tablenotemark{e}  & 1428 & $ 1.41 \pm 0.68 $ & 0.0029 & 0.0258 &$ 0.326 \pm 0.304$&$ 8.64 \pm 0.42 $ & $ 15.2 \pm 7.9$ \\
\tableline{\rule{0mm}{3mm}}
\end{tabular}}
\tablenotetext{a}{low ionization broad absorption line quasars}
\tablenotetext{b}{Non-low ionization broad absorption line quasars}
\tablenotetext{c}{ergs/s}
\tablenotetext{d}{The O III line strength
of the sample was cutoff from above to limit the sample to $< 10^{43}
\mathrm{ergs/s}$}
 \tablenotetext{e}{The O III line strength
of the sample was cutoff from above to limit the sample to $<8 \times 10^{42}
\mathrm{ergs/s}$}

\end{table}

\par In Table 1, we list the physical parameters of various
subsamples for the sake for comparison. Column 1 describes the types
of sources, LoBALQSO or nonLoBALQSO, that comprise the sample. The
second column gives the low end cutoff in $L_{optical}$ that was
used to define the sample. The next column is the number of sources
in the sample. The fourth column is the resultant average
$L_{optical}$. The next two columns give the probability that the
sources within the sample with FWHM $<6500$ km/s are drawn from the
same population as the LoBALQSOs with FWHM $<6500$ km/s, for a
Wilcoxon rank sum test and a K-S test. Columns 7 and 8 are the
Eddington ratio and logarithm of the black hole mass, the derivation
of which are given is the following paragraphs.
\par The sample called, "LoBALQSO de-reddened", is designed to
compare the intrinsic luminosity of the LoBALQSOs to that of the
nonLoBALQSOs. The reddening in the optical is small as noted above
in point 2. Thus, our attempts to de-redden the sources are not
likely to generate large errors, even if our assumptions are not
highly accurate. Our main assumption is based on the DR5 composite
spectra for LoBALQSOs and nonLoBALQSOs in \citet{zha10}. In i-band,
it was found that the LoBALQSO composite continuum had about 0.25 to
0.33 magnitudes of attenuation relative to the nonLoBALQSO composite
continuum. The wavelength of interest in the rest frame, associated
with $L_{optical}$, straddles the high end of i-band and the low end
of the less attenuated z-band. Thus, we expect that an average
extinction of $\approx 0.25$ magnitudes of attenuation in the rest
frame at $5100 \AA$ is a good approximation. The observed flux is
the sum of the reddened AGN spectrum and the contribution of the
host galactic starlight. Therefore, on average, we acknowledge that
the rest frame $L_{opttical}$ computed from observed fluxes will be
larger than it would have been if it were computed directly from the
reddened continuum flux from the nucleus. The 0.25 magnitudes of
attenuation is therefore a liberal upper limit to the average value
of the attenuation and the intrinsic AGN continuum flux will tend to
be over-predicted from the rest frame $L_{opttical}$ computed from
observed fluxes. As such, our estimates will tend to be an upper
limit to the intrinsic average $L_{opttical}$ of the LoBALQSOs. We
used this number to de-redden $L_{optical}$: this amounts to
multiplying each $L_{optical}$ in the LoBALQSO sample by 1.25 to get
the intrinsic value of $L_{optical}$ in the de-reddened sample.
\par It is also of interest to compute the distributions of derived physical properties.
First of all we estimate $M_{bh}$ from the H$\beta$ FWHM using the
virial mass formula in equation (5) of \citet{ves06},
\begin{eqnarray}
 && M_{bh}(H \beta)= \nonumber \\
 && 10^{6.91 \pm
0.02} \left[\left(\frac{F(H\beta)}{1000
\mathrm{km/s}}\right)^{2}\left(\frac{\lambda L_{\lambda}(5100
\AA)}{10^{44}\mathrm{ergs/s}}\right)^{0.50} \right]\;.
\end {eqnarray}

The logarithm of the values are tabulated in the eighth column of
Table 1. The de-reddened sample uses the de-reddened $L_{optical}$
in the formula.
\par The physical parameter of most relevance in certain previous studies is the Eddington
ratio, $\Gamma$ \citep{bor02,sul06}. The interpretation of
\citet{bor02} was largely driven by a principal component analysis
of line properties. The BALQSOs tend to have weak, narrow H$\beta$ emission lines,
weak OIII lines and strong optical FeII lines, giving them low
eigenvector 1 values. The claim is that a low value of eigenvector 1
is physical in origin and is driven by large a Eddington ratio
\citep{bor02}. The corresponding explanation of the narrow H$\beta$
excess that is seen in LoBALQSOs would be that these sources are
very high Eddington ratio sources and this translates into a
stronger ionizing flux which creates low ionization gas farther out
in the gravitational potential (in units of gravitational radius),
hence producing narrower H$\beta$.
\par The first step in
the computation of $\Gamma$ is to estimate $L_{bol}$ from
$L_{optical}$. Perhaps the most popular bolometric correction is
that of \citet{kas00}, $L_{bol}= 9 \lambda L_{\lambda}(5100)$. We
note that this bolometric correction is predicated on $L_{optical}$
representing the intrinsic optical luminosity. This is not the case
for the LoBALQSOs. Hence, the need for the de-reddened sample. The
resultant $\Gamma$ values are tabulated in the seventh
column of Table 1. Since this parameter has received much attention
in the past, in the context of LoBALQSOs, we look at a particular
matched subsample of nonLoBALQSOs in detail. The sample with a
cutoff at $9 \times 10^{44}$ ergs/sec is well matched to the
de-redden distribution of $\Gamma$ according to Table 1. We plot the
distribution of $\Gamma$ in Figure 2 for both the matched sample and the
de-reddened sample. The distributions are indistinguishable in a
statistical sense. The probability that the two samples are drawn
from different distributions of $\Gamma$ is only 0.249 according to
a Wilcoxon rank sum test and 0.254 in a K-S test.

\begin{figure}
\includegraphics[width=160 mm, angle= 0]{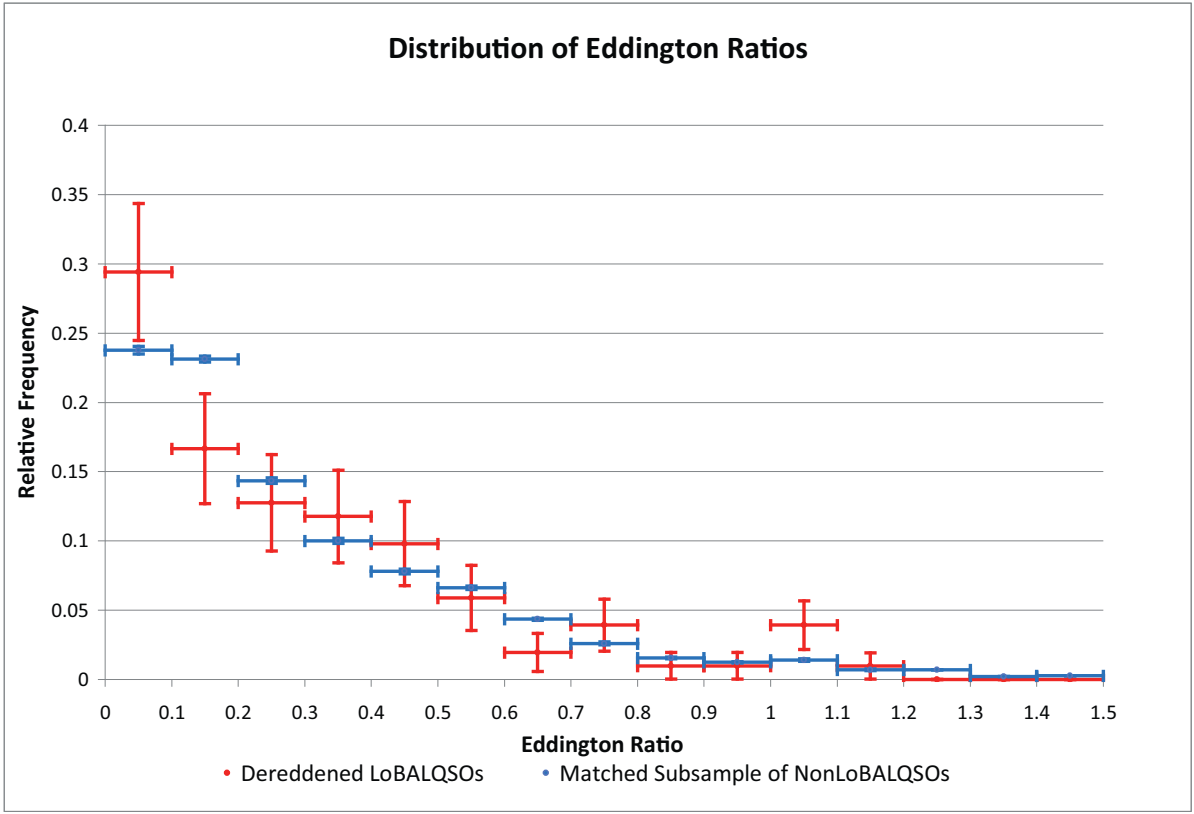}
\caption{Comparison of the distributions of $\Gamma$ for LoBALQSOs
and the "matched" sample of non-LoBALQSOs from DR7. The matched
sub-sample is created from the full SDSS nonLoBALQSOS sample by the
condition $L_{optical}> 9 \times 10^{44}$ ergs/s. The error bars are
based on binomial statistics.}
\end{figure}

Next, we plot the distribution of H$\beta$ FWHM for the matched
sample and the LoBALQSOs (reddening does not affect the LoBALQSO
FWHM distribution) in Figure 3.
\begin{figure}
\includegraphics[width=160 mm, angle= 0]{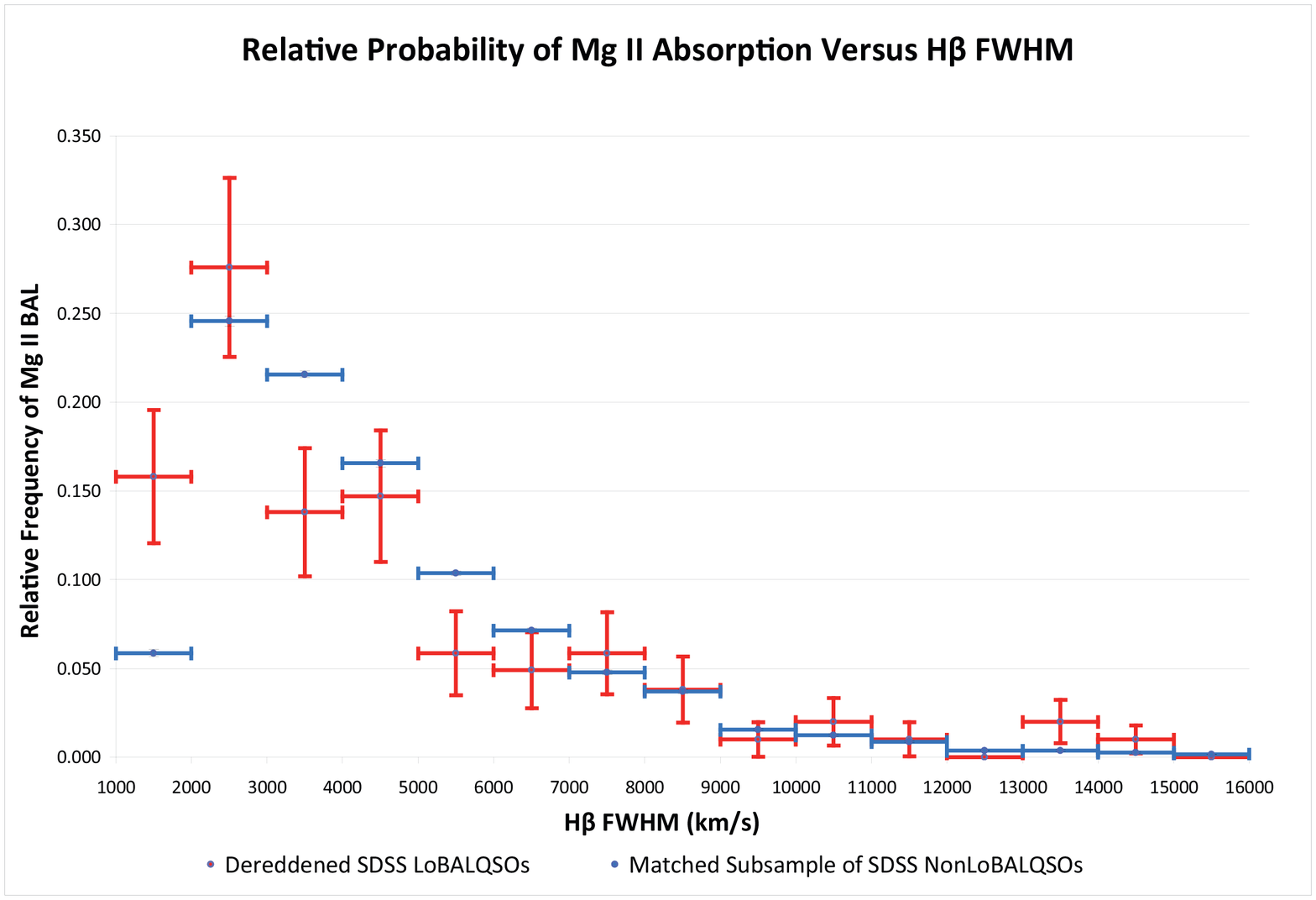}
\caption{The distributions of H$\beta$ FWHM of LoBALQSOs and
"matched" sample of non-LoBALQSOs from DR7. The matched sub-sample
is created from the full SDSS nonLoBALQSOS sample by the condition
$L_{optical}> 9 \times 10^{44}$ ergs/s. Comparison to Figure 1 shows
that the excess of narrow line profiles less than 2,500 km/s in the
LoBALQSO population persists. The error bars are based on binomial
statistics.}
\end{figure}
The distribution is very similar to Figure 1. We conclude that $\Gamma$
is a negligible physical factor with regards
to producing the excess of narrow H$\beta$ profiles for LoBALQSOs in
the Figure 1 and the LoBALQSOs in DR7 in general.
\par In terms of the
eigenvector space, \citet{bor02} claimed that BALQSOs are associated
with very small eigenvector 2 (large accretion rate) and extremely
small eigenvector 1 (large $\Gamma$). Table 1, shows that the
matched subsample in Figures 2 and 3 has a slightly higher average accretion
rate (using the intrinsic $L_{optical}$ as a surrogate) and slightly
higher $\Gamma$ than the de-reddened LoBALQSOs. There are other
samples in Table 1 that have slightly smaller values of
$L_{optical}$ and/or $\Gamma$ as well. For all these samples that
surround the de-reddened LoBALQSOs in the 2-D parameter space, the
excess of narrow lines in the de-reddened LoBALQSO sample persists
with high statistical significance. Thus, there does not appear to
be any physical properties associated with the location in the
eigenvector parameter space that are responsible for the excess of
narrow H$\beta$ line widths in LoBALQSOs.
\par Finally, we explore the relationship to eigenvector 1 in more
detail. In the last column of Table 1, we list the O III EW which
has a large projection on eigenvector 1 \citep{bor92}. Weak O III
emission, in of itself, is not a plausible physical mechanism for
launching a BALQSO wind, but it is possibly related to the BALQSO
phenomenon on the basis of the previously quoted eigenvector space
analysis (i.e., being very far from the accretion disk, the narrow
line region is not a fundamental physical quantity that
characterizes the black hole accretion system such as black hole
mass, Eddington rate and bolometric luminosity). Exploring the
luminosity of the O III emission can potentially provide insight
into the interplay between eigenvector 1 and the existence of
observed UV broad absorption lines. The LoBALQSOs in Table 1 have
weak O III line strengths. None of the control samples that are
created by means of a low end cutoff on $L_{optical}$ have an
average O III EW as small the LoBALQSO sample. We would like the
control smaples to straddle the average value of the OIII EW in the
LoBALQSO sample in order to elucidate the connection between the
H$\beta$ FWHM and the O III EW. Thus, we constructed two modified
samples in the last two rows of Table 1, that are derived from the
$L_{optical}
>9 \times 10^{44}$ ergs/s sample with the additional constraint that the
O III luminosity is bounded from above by $L_{O III} < 10^{43}$
ergs/s or $L_{O III} < 8 \times 10^{42}$ ergs/s. A statistically
significant correlation exists - although the strength of the
correlation seems somewhat diminished relative to the sample (two
rows above in Table 1) without a threshold on $L_{O III}$. There is
a trend in Table 1, as the $L_{O III}$ maximum threshold is lowered
and the average $L_{O III}$ decreases in the three samples defined
by $L_{optical}
>9 \times 10^{44}$ ergs/s, the difference between the partial distribution of H$\beta$
FWHM and the partial distribution of H$\beta$ FWHM in the LoBALQSO
sample becomes less statistically significant. This might be
expected on the basis of the correlation between H$\beta$ FWHM and
$L_{O III}$ in quasar spectra \citep{bor92}. However, the relevant conclusion
drawn from these three samples in Table 1 is that the excess of
narrow H$\beta$ line profiles seen in LoBALQSOs is statistically
significant more pronounced than would have been expected solely
from the consideration of the correlation of $L_{O III}$ with
H$\beta$.

\par In summary, we constructed samples of nonLoBALQSOs from SDSS DR7 that straddled the
LoBALQSO SDSS DR7 sample in three physical properties, optical
luminosity, central black hole mass and Eddington ratio. According
to Table 1, regardless of whether the parameters were above or below
those of the LoBALQSOs, there was always a highly statistically
significant excess of narrow H$\beta$ line profiles for LoBALQSOs.
This was shown explicitly for a sample that was well matched in
Eddington ratio. It is concluded that these parameters are not the
physical origin of the excess narrow line profiles. This conclusion
is not an artifact of reddening, as we verified this for de-redden
LoBALQSOs. This further motivates our exploration of line of sight
effects to produce the excess narrow lines seen in the DR7 LoBALQSO
population.

\section{Simulating a Preferred Line of Sight}In this section, we
consider the distribution of non-LoBALQSO FWHM and simulate what the
distribution would appear to be for different preferred ranges of
the line of sight. This analysis will allow us to discuss the
different orientation dependent models. Anticipating the use of the
simulation to check "goodness of fit" of the binned data with a
$\chi^{2}$ test, we want to minimize the number of free parameters
in our model of the data. The purpose of this fit is to allow us to
model the majority of the sources below 8,000 km/s which comprise
$92\%$ of the entire sample. We consider the $8\%$ of sources with
larger FWHM as outliers that do not drive the orientation dependent
properties of the other $92\%$. As such, we seek a simple single
parameter fit that captures the skewness, peak and width of the
distribution of FWHM below 8, 000 km/sec. A distribution with the
desired skewness and exponential type tail is the Gamma distribution
that we parameterize as,
\begin{eqnarray}
&&
f(\mathrm{FWHM})=N(\mathrm{FWHM})^{-(\alpha-1)}[\mathrm{e}^{-\mathrm{FWHM}/V}]\;,
\end{eqnarray}
where $f(\mathrm{FWHM})$ is the probability density, N is a
normalization constant chosen to make the total probability equal to
one and $V$ is a free parameter that is in units of km/s. We
eliminate the other free parameter, $\alpha$, by choosing which fit
minimizes the $\chi^{2}$ residuals. The result is indicated visually
in Figure 4, the $\alpha = 5 $ plot is a more accurate fit than the
$\alpha = 4$ plot for the best fit values of $V$. This discriminant
is driven by the large residuals generated at low FWHM for $\alpha =
4$. Fine tuning the parameters does not change the fit
significantly.
\begin{figure}
\includegraphics[width=160 mm, angle= 0]{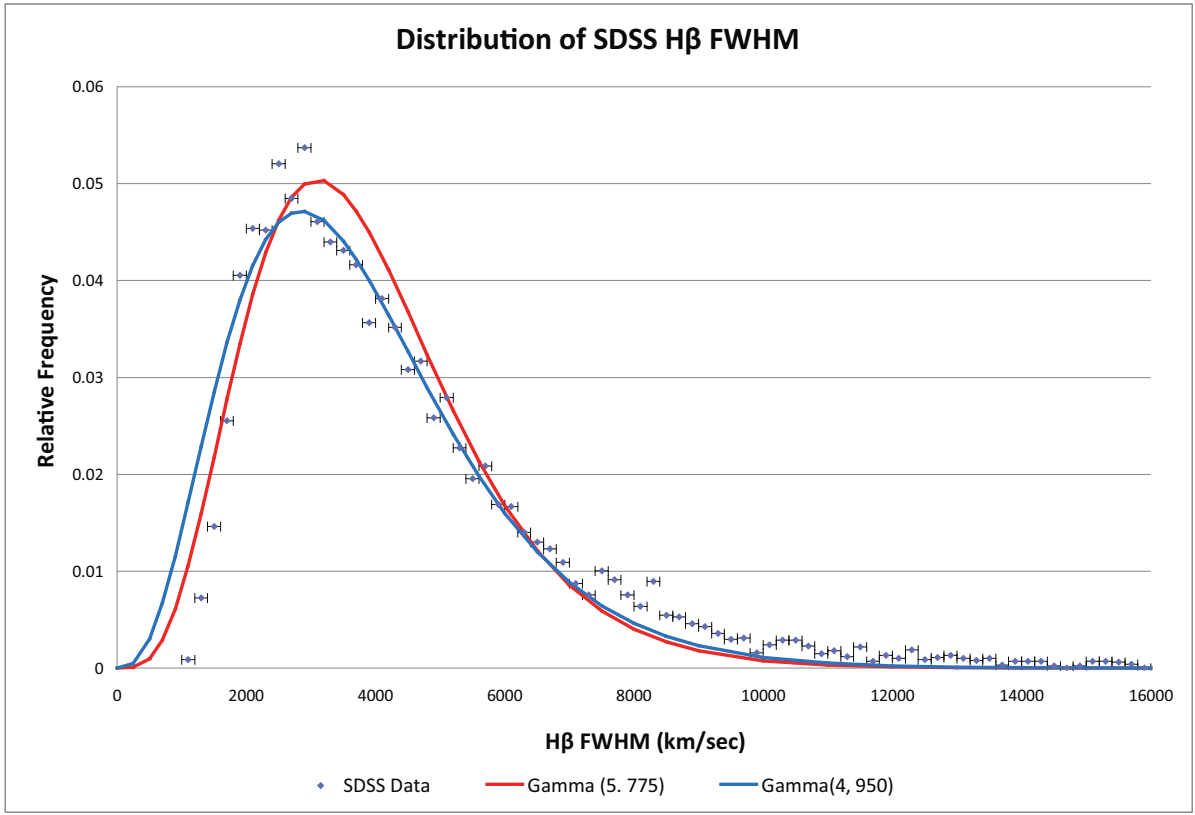}
\caption{Comparison of the theoretical Gamma distributions of
H$\beta$ FWHM of non-LoBALQSOs from DR7 for GAMMA($\alpha=4$,
$V=950$ km/sec) and GAMMA($\alpha=5$, $V=775$ km/sec). The
distribution, GAMMA($\alpha=5$, $V=775$ km/sec), provides a much
better fit to the sharply rising low side of the peak of the
distribution.}
\end{figure}
No simple parametric distribution will fit the data in Figure 4 with
high statistical probability in a $\chi^{2}$ goodness of fit test.
However, as stated above, this theoretical one parameter
distribution with $\alpha = 5 $ and $V$ a function of the
distribution of the line of sight is adequate for estimating the
gross properties of the skewness, variance and mean of the
distribution that are induced by line of sight effects with the
exception of the few percent of outliers residing at high FWHM.
\par We note that the fit for the full sample of nonLoBALQSOs, GAMMA($\alpha=5$, $V=775$ km/sec),
works equally well for the matched subsample that was described in
the last section. Figure 5 shows that the fit to the matched
Eddington ratio subsample of nonLoBALQSOs from section 3 is
very similar to that of the full sample of nonLoBALQSOs. The
figure clearly shows very little difference between the bin
populations of the matched and full samples. There is more scatter
in the matched sample than for the full sample because there are
less than 19\% as many sources in the matched Eddington ratio
sample.
\begin{figure}
\includegraphics[width=160 mm, angle= 0]{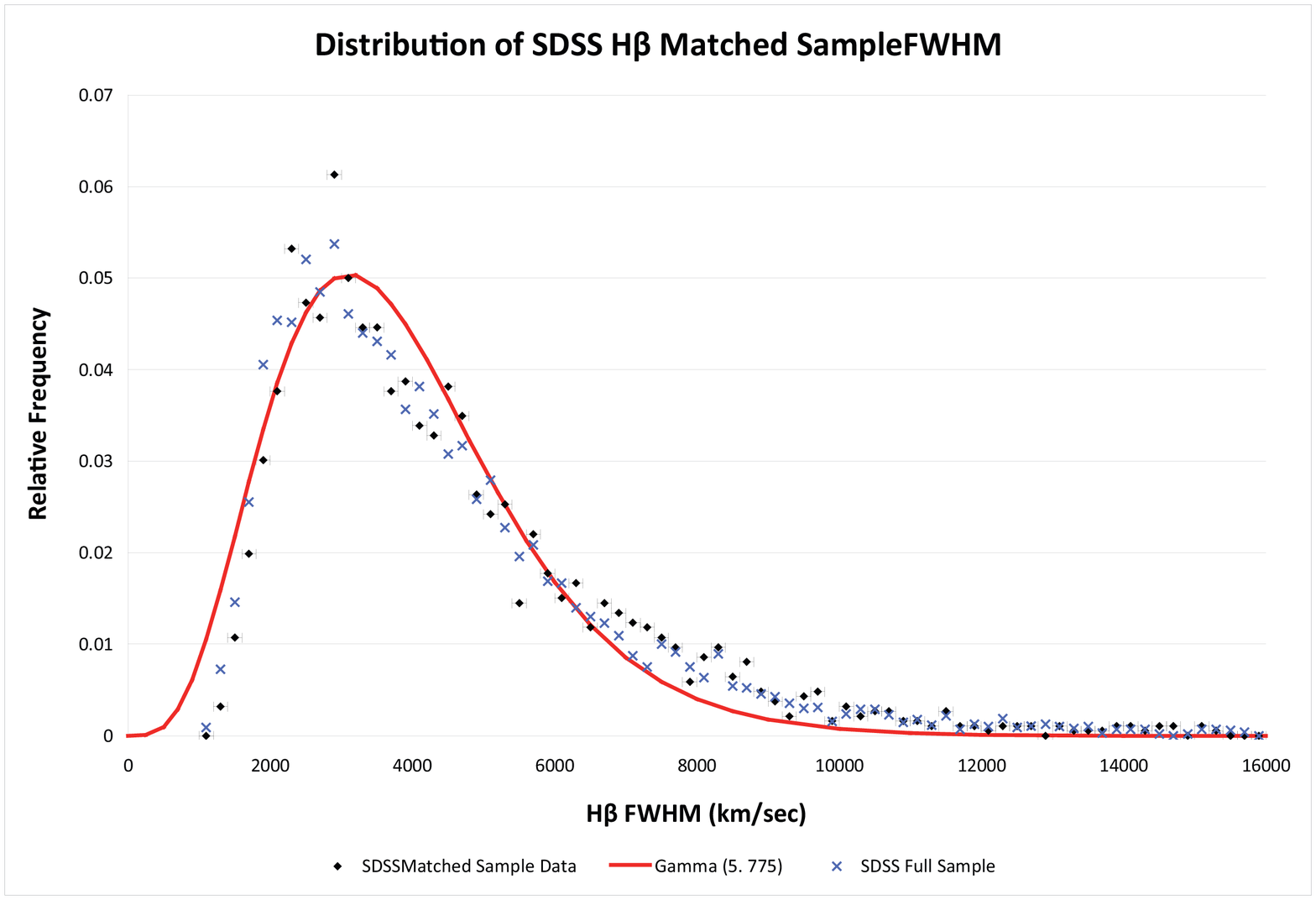}
\caption{Comparison of the theoretical Gamma distribution of
H$\beta$ FWHM of non-LoBALQSOs,GAMMA($\alpha=5$, $V=775$ km/sec)
from Figure 4 and the distribution of H$\beta$ FWHM of the matched
subample of nonLoBALQSOs. The matched sub-sample is created from the
full SDSS nonLoBALQSOS sample by the condition $L_{optical}> 9
\times 10^{44}$ ergs/s. We also show the similarity of the
distribution of the full sample of DR7 nonLoBALQSOs and the matched
sample.}
\end{figure}
\par In order to simulate the change in the distribution of FWHM of DR7
QSOs in Figure 4, if a preferred line of sight is specified, we
consider the popular notion that the BEL gas is distributed in an
equatorial pancake, orthogonal to the jet axis, with a random
velocity, $v_{r}$, superimposed on an equatorial velocity, $v_{p}$,
that is predominantly bulk motion from Keplerian rotation
\citet{jar06,bro86},
\begin{eqnarray}
&& FWHM \simeq 2\sqrt{v_{r}^{2}+v_{p}^{2}\sin^{2}{\theta}}\;.
\end{eqnarray}
In \citet{bro86}, it is suggested that $(v_{r}/v_{p})^2 \approx
0.1$. In general there must be a huge range of these ratios in QSOs,
but the \citet{bro86} value represents a characteristic average
value over the many QSO environments. Thus, we parameterize equation
3 as
\begin{eqnarray}
&& FWHM \approx \sqrt{0.1 + sin^{2}{\theta}}\,v_{blr}\;,
\end{eqnarray}
where $v_{blr}$ is a characteristic dispersion velocity of the BLR.
The analytic forms in equations (2) and (4), allow us to transform
the theoretical $\alpha = 5$ distribution in Figure 4, to different
specified ranges of the line of sight. First, we invoke the results
of \citet{bar90} that are the basis of the "standard model of QSOs"
in \citet{ant93}, QSOs are viewed at a random distribution of lines
of sight within a range of about $45^{\circ}$ (which represents the
half opening angle of the molecular torus) of the jet axis (the
normal to the accretion disk). The transformed distribution is
determined by a change of coordinates defined by the mapping of the
distribution of velocities, $v_{blr}$ to a specified line of sight
by means of equation (4). This amounts to a change in the single
parameter $V$ in equation (2) (and the induced change in probability
normalization, $N$). The distribution of $v_{blr}$ is defined by the
parameter $V=1314\, \mathrm{km/s}\equiv V_{BLR}$. For a given
distribution of line of sights, $g(\theta)$, the distribution
parameter $V$ transforms as $V = F V_{BLR}$, where
\begin{eqnarray}
&& F = \frac{\int{g(\theta)\sqrt{0.1 + sin^{2}{\theta}}\,
\sin{\theta}d\, \theta}}{\int{g(\theta)\, \sin{\theta}d\,
\theta}}\;.
\end{eqnarray}
For the uniform distributions (in terms of solid angle) considered
below, $g(\theta)$ is just a step function. The results are
presented in Figure 6.

\begin{figure}
\includegraphics[width=160 mm, angle= 0]{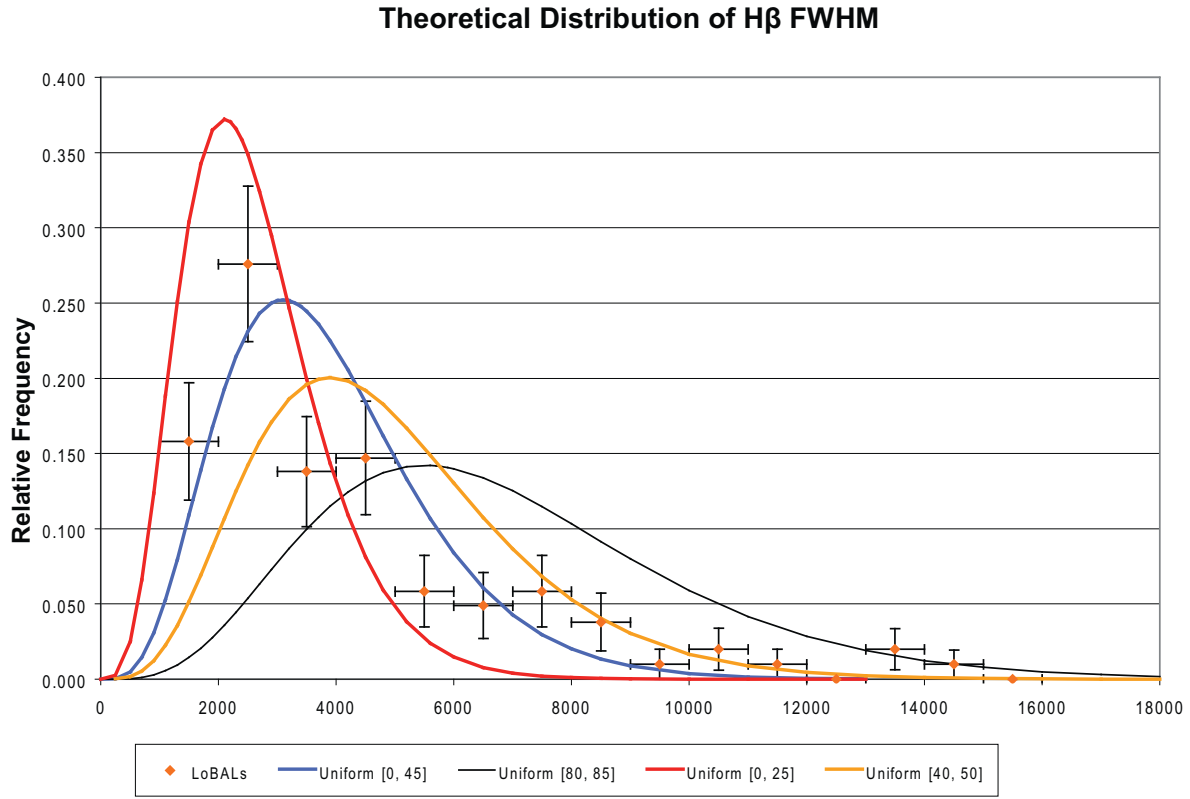}
\caption{Comparison of the simulated distributions of H$\beta$ FWHM
for different distributions of the line of sight. Note that none of
these represent the DR7 LoBALQSO data very well. As the line of
sight becomes more polar the distribution is skewed towards smaller
FWHM.}
\end{figure}
\par In Figure 6, the distributions have the following significance,

\begin{itemize}
\item The uniform distribution of line of sight angles, $0^{\circ}< \theta< 45^{\circ}$, designated as "Uniform [0, 45]" is the \citet{bar90} empirical distribution
for all QSOs
\item The uniform distribution of line of sight angles, $0^{\circ}< \theta< 25^{\circ}$, designated as "Uniform [0, 25]" is the \citet{pun00} theoretical distribution
for LoBALQSOs
\item The uniform distribution of line of sight angles, $80^{\circ}< \theta< 85^{\circ}$, designated as
"Uniform [80, 85]" is the \citet{mur95} theoretical distribution for
LoBALQSOs
\item The uniform distribution of line of sight angles, $40^{\circ}< \theta< 50^{\circ}$, designated as "Uniform [40, 50]"
is the \citet{elv00} phenomenological distribution for BALQSOs
\end{itemize}
We discuss these simulated distributions in the context of the DR7
LoBALQSO FWHM distribution in the next section.
\section{Analysis of the Simulated Distributions} In Table 2, we
attempt to quantify how well the different theoretical models
describe the data. The columns from left to right are first the
model description, then the next two columns are the degrees of
freedom and the $\chi^{2}$ statistic if the continuous distribution
is binned in 1,000 km/s cells as in Figure 1. In order to make a
$\chi^{2}$ analysis plausible, we need a theoretically expected
population of at least 5 per bin. Thus, we combine all the LoBALQSOs
with a FWHM $>$ 7,000 km/s into one final bin (a total of 7 bins are
combined). The fourth column is the probability, $P_{\chi^{2}}$,
that we can reject the hypothesis that the data is described by the
theoretical distribution. Note that we have abbreviated the "Uniform"
Distribution by the symbol "U" in Table 2. We added the K-S "D"
statistic in the fifth column which is a goodness of fit measure
that is not sensitive to small bin populations. The K-S probability
that we can reject the hypothesis that the data is described by the
theoretical distribution is listed in the last column.
\begin{table}
\caption{Goodness of Fit to the Line of Sight Models}
{\footnotesize\begin{tabular}{cccccc} \tableline \rule{0mm}{3mm}
Model  & df& $\chi^{2}$ & $P_{\chi^{2}}$ & D & $P_{KS}$ \\
\tableline \rule{0mm}{3mm}
U [0, 45] &  6 & 11.85 & 0.934 &0.141 &0.97 \\
U [40, 50] &  6&  45.12  &  $>0.9999$& 0.235 &$>0.999$\\
U [80, 85] &  6 &  226.79  &  $>0.9999$& 0.433 &$>0.999$\\
(0.75)(U [0, 25]) + (0.25)(U [80, 85]) & 4 &  3.24 & 0.481& 0.118& 0.87\\
(0.70)(U [0, 25]) + (0.30)(U [80, 85]) & 4 &  3.06 & 0.452& 0.086& 0.71\\
(0.67)(U [0, 25]) + (0.33)(U [80, 85]) & 4 &  3.20 & 0.473& 0.076& 0.66\\
(0.60)(U [0, 25]) + (0.40)(U [80, 85]) & 4 &  4.30 & 0.633& 0.064& 0.60\\
(0.55)(U [0, 25]) + (0.45)(U [80, 85]) & 4 &  5.81 & 0.876& 0.088&0.73 \\
(0.55)(U [0, 25]) + (0.45)(U [40, 50]) & 4 &  4.40 & 0.645 & 0.130& 0.93\\

\tableline{\rule{0mm}{3mm}}
\end{tabular}}
\end{table}

The only fits that are not rejected by one or both of the
statistical tests are the two component fits in rows 4 through 10 (see
Figure 7). Any composite fit that is composed of the range 55\%
Uniform [0, 25] and 45\% Uniform [80, 85] to 75\% Uniform [0, 25]
and 25\% Uniform [80, 85] can not be rejected on a statistical
basis. Note that the equatorial model is rejected by this
statistical analysis.
\begin{figure}
\includegraphics[width=160 mm, angle= 0]{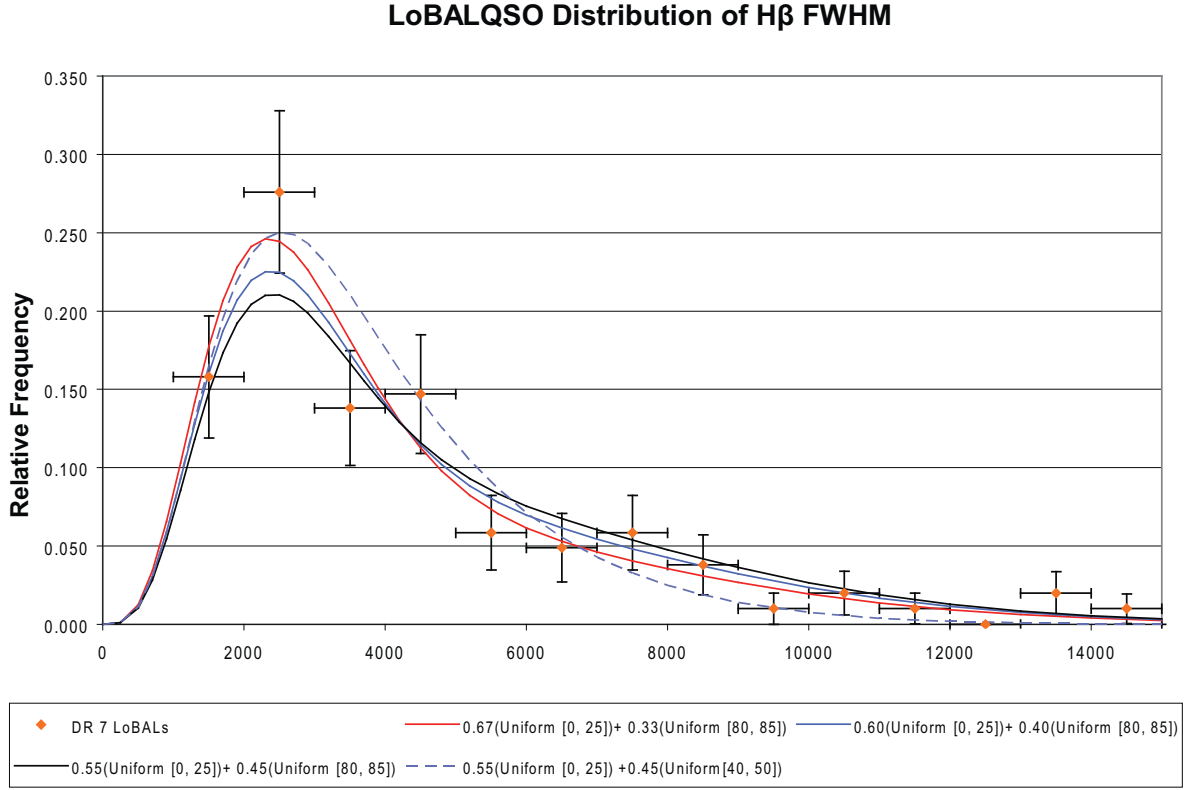}
\caption{Comparison of the simulated two component distributions of
H$\beta$ FWHM for lines of sight defined by, (0.67)(Uniform [0, 25])
+ (0.33)(Uniform [80, 85]), (0.60)(Uniform [0, 25]) + (0.40)(Uniform
[80, 85]), (0.55)(Uniform [0, 25]) + (0.45)(Uniform [80, 85]) and
(0.55)(Uniform [0, 25]) + (0.45)(Uniform [40, 50]). The model
(0.67)(Uniform [0, 25]) + (0.33)(Uniform [80, 85]) is tantamount to
notion that 2/3 of the LoBALQSOs are polar wind sources that were
proposed in the theoretical model of \citet{pun00} and 1/3 of the
LoBALQSos are equatorial winds sources proposed in the theoretical
wind model of \citet{mur95}.}
\end{figure}
The distribution, (0.67)(Uniform [0, 25])+ (0.33)(Uniform [80, 85]),
represents the peak fairly accurately and it is also fits the high
velocity tail reasonably well.

\section{Discussion}
There have been other discussions involving the possible orientation
of the line of sight to the BAL gas which we discuss below. What
distinguishes our methods from previous efforts is the very direct
determination of the geometry. Previous discussions in the
literature involved indirect physical arguments that invariably
relied upon unverifiable assumptions. In this section we discuss our
results in the context of previous discussions of orientation and
line widths.
\subsection{Polarization}
Historically, the observed elevated polarization of BALQSOs has been
used as an argument that ordinary quasars appear as BALQSOs because
the line of sight is just above the dusty torus in the standard
model \citep{sch97,ogl97,lam04}. In axisymmetric geometries,
equatorial lines of sight produce higher degrees of polarization
than polar lines of sight. Furthermore, the putative "equatorial"
line of sight to the BAL gas is consistent with the idea that stray
dusty gas above the torus is responsible for the reddening that is
observed in the UV continuum \citep{sch97}. This argument is
indirect and very circumstantial. First of all, the elevated degree
of polarization is not that high for most BALQSOs, the median
polarization is $\gtrsim 1\%$. Theoretically, this value can be
accommodated in any geometry due to modest attenuation of the
continuum (see below) by the BAL wind which enhances the prominence
of the scattered light \citep{pun00}. Secondly, two of the most
polarized BALQSOs, FIRST 1556+3517, \citet{bro97}, and MRK 231,
\citet{smi95}, have polarizations $\approx 13\%$ in the near UV and
are now know to be polar objects \citep{gho07,rey09}. Furthermore,
both of these polar sources are heavily reddened
\citep{naj00,smi95}. Thus, these two well studied examples show that
one can not use the enhanced polarization of BALQSOs and reddening
of the continuum as an argument for equatorial BAL winds.
\par Theoretically, it was pointed out in
\citet{pun00} that the rare outlier polarizations of $>3\%$ can be
obtained in a polar line of sight if the assumption of axisymmetry
is dropped. Modest reddening is a natural consequence of the polar
wind model. There is significant attenuation from scattering in the
theoretical models as the continuum radiation field propagates
through the polar BAL wind \citep{pun00}. The amount of attenuation
is consistent with the statistically based estimates of
\citet{goo97}, $\sim 30\%$ to $50\%$. The UV must shine through
larger optical depths (the inner regions of the wind) than the
optical (outer regions of the polar wind), hence the polar models
naturally produce reddening. Excessive reddening requires
entrainment of dusty material on larger scales.
\par The equatorial
line of sight argument is very dependent on the assumption of
axisymmetry for the scattering surface, which is apparently not
justified (as noted above, the extreme outlier elevated
polarizations can occur in polar geometries once the perfect
axisymmetry assumption is dropped and evidenced by the high
polarization in the polar sources MRK 231 and FIRST 1556+3517).
Within these equatorial models the LoBALQSOs are viewed even closer
to the equator than other BALQSOs \citep{mur95}. However, the fits
in Table 2 and Figure 6 indicates that this model cannot explain the
DR7 data. The equatorial view enhances large FWHM and cannot explain
the excess of small FWHM seen in the DR7 LoBALQSO data. Figure 7
indicates that at most $\approx 1/3$ of the LoBALQSOs are equatorial
outflows. The method used in this paper to determine the line of
sight requires less physical interpretation of the data and less
assumptions than the polarization and reddening argument.
\subsection{Radio Luminosity and Orientation} The anti-correlation between radio luminosity and the
degree of absorption in LoBALQSOs was used in \citet{xin10} as an
argument that most LoBALQSOs must be viewed through an equatorial
BAL wind. They proposed that the LoBALQSOs tend to be weak radio
sources because the radio emission is vastly weaker for
perpendicular lines of sight to a relativistic jet. They made an
analogy to powerful radio loud sources in which radio cores are
enhanced by Doppler beaming \citep{bro86}. It is far from obvious
that there is a strong connection between radio loud quasar jets and
the weak radio emission seen in most LoBALQSOs. The authors assume
without justification that every LoBALQSO jet is identical (bulk
Lorentz factor, power, etc.) and the difference in single point
radio flux density is due to the line of sight. Core flux is not a
good orientation indicator, since the same type of logical argument
presented in \citet{xin10} leads to the erroneous conclusion that
radio quiet quasars must be viewed near the equator because they are
radio weak (compared to radio loud sources). The true orientation
indicator is the ratio of core to lobe flux density as noted in
\citet{bro86}. The data analysis presented in our paper suggests the
alternative explanation that the mechanism that launches the BAL
wind tends to inhibit the central engine that drives the powerful
radio jet or the ability of that jet to propagate.
\subsection{Two Component Models} The only other
attempt that we know of to arrive at the distribution of lines of
sights to BALQSOs was in \citet{bor10}. They fit CIV absorption
(HiBALQSOs) line profiles in a parametric model. Unfortunately, the
model is completely adhoc, it is comprised of a spherical fast wind
with with a slower narrow equatorial wedge inserted. The bulk of the
solid angle is called a polar wind, yet it is nearly spherical.
Every parameter is adhoc, from the launch point, to the distribution
of the broad emission line gas, the density and the velocity. They
found the need for both components, in general, to reproduce the
line shapes. In \citet{pro04}, the numerical simulations showed that
by adjusting the parameters one could get polar winds, equatorial
winds, mid-latitude winds as in \citet{elv00} or virtually any
combination thereof. However, given the adhoc nature of the models
in \citet{bor10}, it is very unclear if there is any physical
significance to the parametric variations that led these authors to
conclude that most sources are viewed along the equatorial plane.
The driving piece of logic was that deep zero velocity absorption (P
Cygni type) lines mean that one is viewing the BAL wind along the
eqautor. However, FIRST1556+3517 has deep zero velocity absorption
in MgII, \citet{bro97}, and it is polar.

\section{Conclusion}In this paper, we used the distribution of H$\beta$ FWHM in the SDSS
DR7 data release to estimate the distribution of lines of sight to
LoBALQSOs. Our analysis indicates that predominantly equatorial
outflow is ruled out for LoBALQSOs. We also find that the data is
not well represented by random lines of sight. The distribution has
an excess of sources with narrow H$\beta$ that is best fit by
assuming two classes of LoBALQSOs, the majority ($\approx 2/3$) are
polar outflows and the remainder are equatorial outflows. By
choosing a variety of subsamples of nonLoBALQSOs matched in redshift
that straddle the values of luminosity, black hole mass and
Eddington ratio of the de-reddened LoBALQSO sample, we find that the
narrow line excess in the LoBALQSO sample persists in all cases and
therefore eliminate the possibility that the excess narrow lines
seen in LoBALQSOs arise from the physical properties of the central
black hole accretion system such as Eddington ratio.
\par We make one more comment on the important work of
\citet{gan07}. They measured the FWHM of the Mg II of HiBALQSOs and
non-BALQSOs. There was no statistical difference. This implies
random lines of sight to HiBALQSOs per the methods discussed here.
However, they chose to combine mini-BALQSOs and BALQSOs to increase
sample size. These types of sources are not BALQSOs in the
conventional sense and the results might not be reliable. The DR5
statistical analysis of \citet{zha10} indicate that these types of
sources (mini-BALQSOs have a large overlap in definition with the
intermediate width absorption line sources of \citet{zha10}) tend to
resemble non-BALQSOs more than BALQSOs in many spectral properties.
Furthermore, one must use caution in the interpretation of Mg II
FWHM with HiBALQSOs, since there might be low level absorption in
the continuum blueward of Mg II that might skew continuum estimates.
It would be hard to segregate this effect and quantify its impact on
the sample. This type of analysis is best done without resorting to
resonance lines for FWHM estimates. Thus, H$\beta$ is preferred
over Mg II.
\par Based on these considerations mentioned above, it would be important
to reproduce the analysis for LoBALQSOS presented here with a
similar study for Hi BALQSOs. One could select high redshift
HiBALQSOs from SDSS and observe the H$\beta$ profiles in the IR.

\begin{acknowledgements}
We would like to thank Robert Antonucci for sharing his expertise
and insights. We were also fortunate to benefit from many
interesting comments from Paola Marziani. We would also like to
thank an anonymous referee who offered many ideas that improved the
manuscript.
\end{acknowledgements}

\end{document}